\documentstyle[12pt,epsf]{article}
\catcode`\@=11
\def\numberbysection{\@addtoreset{equation}{section}
\def\theequation{\arabic{section}.\arabic{equation}}}
\def\appendix#1{
  \addtocounter{section}{1}
  \setcounter{equation}{0}
  \renewcommand{\thesection}{\Alph{section}}
  \renewcommand{\theequation}{\Alph{section}.\arabic{equation}}
  \section*{Appendix \thesection\protect\indent #1}
  \addcontentsline{toc}{section}{Appendix \thesection\ \ \ #1}
  }
\renewcommand{\thefootnote}{\fnsymbol{footnote}}
\newlength{\extraspace}
\newlength{\extraspaces}
\newcommand{\be}{\begin{equation}
\addtolength{\abovedisplayskip}{\extraspaces}
\addtolength{\belowdisplayskip}{\extraspaces}
\addtolength{\abovedisplayshortskip}{\extraspace}
\addtolength{\belowdisplayshortskip}{\extraspace}}
\newcommand{\ee}{\end{equation}}
\newcommand{\ba}{\begin{eqnarray}
\addtolength{\abovedisplayskip}{\extraspaces}
\addtolength{\belowdisplayskip}{\extraspaces}
\addtolength{\abovedisplayshortskip}{\extraspace}
\addtolength{\belowdisplayshortskip}{\extraspace}}
\newcommand{\ea}{\end{eqnarray}}
\newcommand{\nonu}{\nonumber \\[.5mm]}
\newcommand{\tr}{\, {\rm tr} \,}
\newcommand{\e}{\, {\rm e}}
\newcommand{\A}{\!\!}
\newcommand{\del}{\partial}
\newcommand{\half}{{\textstyle{1\over 2}}}
\newcommand{\ie}{{\it i.e.}\ }
\begin{document}
\addtolength{\baselineskip}{.5mm}
\thispagestyle{empty}
%
\begin{flushright}
OU--HET 208 \\
December, 1994 \\ %
\end{flushright}
\vspace{3mm}
\begin{center}
{\Large{\bf{ Modification of Matrix Models by Square Terms of Scaling
Operators }}} \\[24mm]
{\sc Hiroshi Shirokura}\footnote{A JSPS Research Fellow,\quad
e-mail: siro@funpth.phys.sci.osaka-u.ac.jp} \\[8mm]
{\it Department of Physics, Osaka University \\[3mm]
Toyonaka, Osaka 560, JAPAN} \\[29mm]
%
{\bf ABSTRACT}\\[9mm]
{\parbox{13cm}{\hspace{5mm}
We study one (or two) matrix models modified by terms of the form
$g(\rho(P))^2 + g'(\rho'({\cal{O}}))^2$,
where the matrix representation of the puncture operator $P$ and the one
of a scaling operator ${\cal{O}}$ are denoted by
$\rho(P)$ and $\rho'({\cal{O}})$ respectively.
We rewrite the modified models as effective theories of baby universes.
We find an upper bound for the gravitational dimension of
${\cal{O}}$ under which we can fine tune the coupling constants to obtain
new critical behaviors in the continuum limit.
The simultaneous tuning of $g$ and $g'$ is possible
if the representations $\rho(P)$ and $\rho'({\cal{O}})$ are chosen
so that the non-diagonal elements of the mass matrix of the effective
theory vanish.
}}
\end{center}
\vfill
\newpage
\renewcommand{\thefootnote}{\arabic{footnote}}
\setcounter{section}{0}
\setcounter{equation}{0}
\setcounter{footnote}{0}
\numberbysection
%
%
\section{Introduction}
\label{Intro}

Matrix models \cite{BIPZ} have played an important role in the theory of
quantum gravity in two dimensions.
The existence of the double scaling limit \cite{GRMI} is a feature of
matrix models.
This enables us to extract nonperturbative information from matrix models.
Until now, many types of matrix models have been studied in detail and
people have made every effort to gain new information from
these models.
We know much about the matrix models which are identified to
Liouville theory coupled to the $c\leq 1$ minimal conformal matter.
We hope that we can jump over the barrier of $c=1$ or find new
universality classes in the continuum limit by modifying these conventional
models or creating new models.

Modified matrix models which we study in this paper have contact terms
such as $g(\tr \Phi^4)^2$ in the action.
The contact terms create surfaces with microscopic wormholes.
By changing coupling constants of these square terms, we can change
the weight for surfaces with wormholes in the matrix integral.
This kind of modified matrix models has been studied by many people for a
long time \cite{DDSW}.
An interesting point of these models is the existence of a
critical value $g_t$ for the coupling constant $g$.
For $g < g_t$, the model belongs to the same universality as the
$g = 0$ model.
For $g > g_t$, as easily expected, one cannot take the double scaling
limit any more and degenerate surfaces dominate.
When we fine tune the coupling constant to $g_t$, a new type of critical
behavior appears in a double scaling limit and the string susceptibility
$\gamma$ jumps to a new positive value $\bar{\gamma}$.
At present, similar results are confirmed in the context of conventional
matrix models --- 1-matrix models, 2-matrix models and the $c = 1$ model.
In general, the new exponent $\bar{\gamma}$ is related to the original
negative value $\gamma$ as
\be
\bar{\gamma} = \frac{\gamma}{\gamma-1}.
\label{RelExps}
\ee

Recently, an extension of these modified matrix models is investigated
\cite{KLEHASHI}.
Instead of the square term of the puncture operator,
the authors of Ref. \cite{KLEHASHI} included a term of the form
$g'(\rho'({\cal{O}}))^2$ to the action,
where $\rho'({\cal{O}})$ is a representation of a scaling operator
${\cal{O}}$ by matrices.
They found a critical value for $g'$.
When we fine tune $g'$ to the critical value, the gravitational dimension
of this scaling operator changes to a negative value.

According to our common knowledge of matrix models, the positive
string susceptibility at $g = g_t$ and the negative gravitational dimension
at $g' = g'_t$ are very strange.
Recently, a simple explanation for these phenomena is proposed
in language of Liouville theory \cite{KLEB}.
Let us have an introductory remark for the Liouville theory to explain
the logic clearly.

The path integral over a closed Riemann surface of genus $g$ is given by
the following form
\be
Z_h = \int\A d\tau\A\int\![d\Psi][d\phi][db][dc]
      \e^{-S_\Psi-S_\phi-S_{int}-S_{b,c}},
\label{LiouvillePath}
\ee
where $\phi$ is the Liouville field and $\Psi$ is the $(p, q)$
minimal matter field
\footnote{We assume in this paper that $p$ and $q$ are coprime integer and
$p,q \ge 2$. The central charge of this minimal matter is smaller than $1$.}.
In (\ref{LiouvillePath}) $S_\Psi$ is the action for the matter and
$S_{b,c}$ the action for the ghosts.
The sum of the Liouville action and the interaction term is written as
\ba
\lefteqn{S_\phi+S_{int}} \nonu
& = & \frac{1}{8\pi}\int\A d^2z\sqrt{\hat{g}}
      [\hat{g}^{ab}\del_a\phi\del_b\phi-Q\hat{R}\phi
       +tO_{\rm min}(\Psi)\e^{\alpha\phi}
       +t_{\cal{O}}O(\Psi)\e^{\beta\phi}].
\label{LiouvilleAction}
\ea
In (\ref{LiouvilleAction}), we have two interaction terms.
The coupling constant $t$ of the first interaction term
is the cosmological constant and $O_{\rm min}(\Psi)$ is the
conformal field which has the lowest weight
$h_{\rm min} = \frac{1-(p-q)^2}{4pq}$.
Similarly, $t_{\cal{O}}$ is a coupling constant of the source term of the
scaling operator and $O(\Psi)$ is a conformal field whose weight is $h$.
The parameters $\alpha$ and $\beta$ in (\ref{LiouvilleAction}) are determined
by requiring the puncture operator
$P = \int\A d^2zO_{\rm min}\e^{\alpha\phi}$
and the scaling operator
${\cal{O}} = \int\A d^2zO\e^{\beta\phi}$
should be conformally invariant.
The conditions for the conformal invariance are
\ba
h_{\rm min} - \frac{1}{2}\alpha(\alpha+Q) & = & 1, \nonu
h - \frac{1}{2}\beta(\beta+Q) & = & 1.
\ea
We denote the larger solution and the smaller one by $\alpha_+$ and
$\alpha_-$ respectively.
Similarly, we have two solutions $\beta_+$ and $\beta_-$.
We usually adopt $\alpha_+$ and $\beta_+$ as they have a connection to the
semiclassical limit of the Liouville theory \cite{SEIBERG}.
The string susceptibility and the gravitational dimension of the scaling
operator ${\cal{O}}$ are found by shifting the zero mode of the Liouville
field \cite{DDK},
\ba
\gamma & = & 2 + \frac{Q}{\alpha_+} = - \frac{2}{p+q-1}, \nonu
d      & = & 1 - \frac{\beta_+}{\alpha_+} .
\ea

The proposal of Klebanov and Hashimoto \cite{KLEHASHI} is that
the jump in the string susceptibility and the gravitational dimension
observed in modified theories can be interpreted as a consequence of
the change in the branch of $\alpha$ or $\beta$.
If the branch of $\alpha$ changes from $(+)$ to $(-)$,
the exponent jumps to a new value
\be
\bar{\gamma} = 2 + \frac{Q}{\alpha_-} = \frac{2}{p+q+1}.
\ee
This agrees with the matrix model result (\ref{RelExps}).
This change of the branch changes the gravitational dimension
of the scaling operator as well.
The jump in the gravitational dimension found in \cite{KLEHASHI}
is also understood as a consequence of the change of the branch in $\beta$.

%
%
\subsection{Summary of Results}

In this paper, we study matrix models modified by two square terms
of the puncture operator and a scaling operator
and find possible continuum limit changing the coupling constant of the
square term of the puncture operator $g$ and the one of the scaling
operator $g'$.
Our strategy to this problem is similar to \cite{KLEHASHI}.
We linearize the action by introducing baby universes and
perform the matrix integration first.
We regard the effective action derived by this matrix integration
as the effective action of a theory of baby universes.
We derive the expansion form of the effective action near the critical
point of the unmodified theory.
The free energy is expanded around the saddle point of the effective action
and the integrations over baby universes are performed.
We read scaling characteristics from the result of these integrations.
Figure 1 summarizes the universality of the modified models.
In this figure, the constants ${a_2}^{-1}$ and ${b_2}^{-1}$ depend on
the details of the matrix model.
The structure of the universality diagram of the generic case (Fig.\,1(a))
is different from the special case (Fig.\,1(b)).
As a result we find two new aspects of the modified theory.
The first and the most important discovery is that change of the branch
in $\beta$ never occurs for scaling operators with gravitational
dimension $d > \frac{1}{3}(1+\gamma)$.
This means that we cannot get new critical behavior by introducing
square terms of such scaling operators.
The second is that we must choose the representation of
the scaling operators in matrices carefully to realize simultaneous tuning
of the coupling constants of the modification terms as Fig.\,1(b)
\footnote{Klebanov and Hashimoto discussed the problem of
simultaneous tuning in \cite{KLEHASHI}.
However, the fact that odd point functions of a scaling operator
does not always vanish modifies their results slightly.}.
We confirm that the tuning of the coupling constant means that
the branch of the Liouville dressing $\alpha$ or $\beta$ changes to
the opposite branch.
In language of the theory of the baby universes, this fine tuning means
that the baby universe corresponding to the scaling operator becomes
massless.
Of course the relation between the modified free energy and unmodified
one derived in \cite{KLEHASHI} is valid if we choose the representations
of scaling operators appropriately.
\begin{center}
\leavevmode
\epsfysize=6cm \epsfbox{uni.ps}
\end{center}
\begin{center}
{\parbox{15cm}{
\small
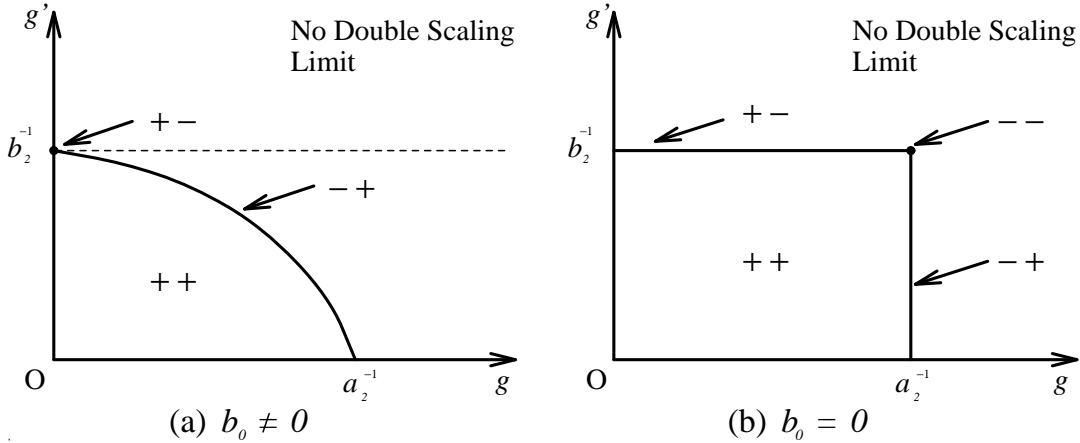
Figure 1: This figure shows the universality of the modified matrix
models when $d\leq\frac{1}{3}(1+\gamma)$.
The universality is represented by the branches of the Liouville
dressings $\alpha$ and $\beta$ like $(+\,-)$.
The left hand side is the branch for $\alpha$ and the right hand side
the one for $\beta$.
When the mass matrix of the effective theory is non-diagonal, the diagram
has no simultaneous tuning point $(-\,-)$ as in Fig.\,1(a).
Fig.\,1(b) shows that the simultaneous tuning is possible only when the mass
matrix becomes diagonal.
}}
\end{center}

%
%
The rest of the paper is organized as follows.
In section 2 we discuss the modified matrix models which contain the square
term of the puncture operator and the one of a scaling operator.
We find the precise form of the analytic part in the free energy of
the unmodified model expanded around a critical point in Appendix A.
Using this, we solve the saddle point equations and find the general
form of the modified free energy.
We read the critical exponents and complete the universality diagram
depicted in Fig.\,1.
In section 3 we discuss some problems of the representation of scaling
operators in the modified matrix models.
We summarize the results and discuss in section 4.
In Appendix A we perform matrix integration of the 1-matrix model with
potential of the sixth order to clarify the generic form for the non-singular
part of the free energy of the conventional matrix models.

%
\section{Matrix Models Modified by Square Terms of Scaling Operators}
\label{MMMod}

Matrix models modified by square terms of
general scaling operators has studied by Klebanov and Hashimoto.
The Ising model coupled to gravity and the generalization of it have
been studied by them.
They assumed that odd-point functions of a scaling operator must vanish.
In general, this is not always true.
This point has a serious influence on the universality of the modified
models.
Their argument must be modified because of their inexact assumption.
Therefore we will study the modified matrix models carefully.

In general the explicit form of matrix integral depends on the details
of the model.
Then we give the modified matrix models in a general way
and study the universality of the models completely.
The definition of matrix models modified by square terms is simply
given by
\[
Z = \int\A D\Phi\,\e^{-NS},
\]
\be
S = S_{\rm cr} + \Delta_0\rho(P)
    + \sum_{i=1}^n(\tau_i)_0\rho_i({\cal{O}}_i) \nonu
    -\frac{g}{2N}(\rho(P))^2
    -\sum_{i=1}^m\frac{g_i}{2N}(\rho_i({\cal{O}}_i))^2,
\label{ModifiedMM}
\ee
where we denote the matrix measure symbolically by $D\Phi$.
For 1-matrix models this is simply the sum over a matrix $\Phi$
and for 2-matrix models it means the sum over two matrices.
In the action, $S_{\rm cr}$ is the fine tuned action of the matrix model
which realizes Liouville theory coupled to $(p,q)$ minimal conformal
matter in the continuum limit
\footnote{We exclude the $c=1$ model here.}.
$\Delta_0$ is the bare cosmological constant.
We consider the cosmological constant as the coupling constant for the
puncture operator $P$.
$(\tau_i)_0$ are the bare coupling constants for the scaling operators
${\cal{O}}_i$ which have the gravitational dimension $d_i$ in the
unmodified model.
We assume that none of the scaling operators have the same gravitational
dimension.
We express the matrix representations of the puncture operator
and the scaling operators ${\cal{O}}_i$ as $\rho(P)$ and
$\rho_i({\cal{O}}_i)$ respectively.
The coupling constants of the square terms $g$ and $g_i$ and
the choice of the representations $\rho$ and $\rho_i$ characterizes
the modified models.
Varying these parameters we study the universality classes of our models
in the continuum limit.
We assume that the number of the source terms of scaling operators
must be larger than the number of the square terms of scaling operators,
\ie, $n \geq m$.
We need a source term in the action for each square term.

{}From our definition of the modified matrix models, the jump of the
string susceptibility has been observed in the case where only the square
term of the puncture operator exists..
In this paper, we mainly discuss the case $n=m=1$.
Essential points of our models appear in this simple case.
The action of the modified model in this case is written as
\be
S = S_{\rm cr} + \Delta_0\rho(P)+\tau_0\rho'({\cal{O}})
    -\frac{g}{2N}(\rho(P))^2-\frac{g'}{2N}(\rho'({\cal{O}}))^2 .
\label{DefOfMMM}
\ee

As the action $S$ contains square terms we cannot perform the matrix
integration directly.
Applying a trick, we can rewrite (\ref{ModifiedMM}) in the
following form
\be
Z(\Delta_0,\tau_0,g,g')
= \frac{N^2}{2\pi\sqrt{gg'}}\int\A dy\,\e^{-\frac{N^2gy^2}{2g}}
  \int\A dv\,\e^{-\frac{N^2v^2}{2g'}}
  \int\A D\Phi\,\e^{-N\tilde{S}} ,
\label{LinearizedForm}
\ee
\be
\widetilde{S} = S_{\rm cr}
                + (\Delta_0-y)\rho(P) + (\tau_0-v)\rho'({\cal{O}}) .
\ee
At this stage we can perform the matrix integration in
(\ref{LinearizedForm}).
It is important to understand qualitative structure of
the logarithm of the matrix integral $\int\!\!D\Phi\,\e^{-N\tilde{S}}$
expanded around the critical point of the unmodified model.
We discuss this in appendix A and the result is given by
\[
\log\,\int\!\!D\Phi\,\e^{-N\tilde{S}} =
\mbox{Analytic Part}\,+ F(x,u,N^2),
\]
\ba
F(x,u,N^2)
& = & N^2\left(-\frac{a_3}{(2-\gamma)(1-\gamma)}
      x^{2-\gamma} + \cdots\right)
      - N^0(a_4\log x + \cdots) \nonu
&   & - N^{-2}(a_5x^{-2+\gamma} + \cdots) + {\cal{O}}(N^{-4}),\nonu
&   & - N^2u(b_3x^{d+1-\gamma} + \cdots)
      - \half N^2u^2(b_4x^{2d-\gamma}
        + \cdots) \nonu
&   & - {\textstyle{1\over 6}}N^2u^3(b_5x^{3d-1-\gamma} + \cdots)
      + \cdots + {\cal{O}}(N^0ux^{d-1}) .
\label{ScalingForm}
\ea
It is necessary for us to know an explicit form of the analytic
part of this expansion.
The structure depends on the value of the gravitational dimension of the
scaling operator ${\cal{O}}$.
On the other hand, the structure of the most singular part
(\ref{ScalingForm}) holds for any scaling operators.
In general the analytic part of the free energy contains contributions
from the higher-genus surfaces.
If we restrict the upper limit of the gravitational dimension of the
scaling operator to $1$, the analytic part in $\Delta$ and $\tau$ comes
from only planar graphs.
Besides, when we limit to $d\leq\frac{1}{3}(1+\gamma)$,
the non-singular part contains at most bilinear terms with respect to
$x$ and $u$,
\ba
\log Z(x, u)
& = & N^2(-a_1x + \half a_2x^2) + N^2(-b_1u + \half b_2u^2)
      - N^2b_0xu \nonu
&   & + F(x,u,N^2).
\ea
When we study modified matrix models, it is essential
to distinguish the case of $d\leq\frac{1}{3}(1+\gamma)$ from
the case of $d > \frac{1}{3}(1+\gamma)$.
As far as we consider conventional matrix models, the condition
$3d - 1 - \gamma \leq 0$ leads $0< 2d - \gamma < 1$ and
$1 < d + 1 - \gamma < 2$.
Thus the above form of the non-singular part is guaranteed only by the
condition $d \leq \frac{1}{3}(1+\gamma)$.
When $d > \frac{1}{3}(1+\gamma)$, some terms such as $\Delta\tau^2$ appear
in the bulk part.
As far as we consider the conventional matrix models,
these terms are not important since we discard the analytic terms
when we take the double scaling limit.
If we don't fine tune the coupling constants, these terms is irrelevant
even in the modified theory.
If we fine tune them, however, such terms cause scaling violation:
we cannot take any double scaling limit.
We will discuss this point later.
For a while we concentrate on scaling operators which satisfy
the restriction $d \leq \frac{1}{3}(1+\gamma)$.

Rewriting the integration variables in terms of
$x = \Delta_0 - y,\quad u= \tau_0 - v$, we find
\[
\log Z(\Delta_0,\tau_0,g,g') =
\log N^2\int_{-\infty}^\infty\A dx \int_{-\infty}^\infty\A du\,
\e^{f(x,u)} ,
\]
\ba
f(x,u) & = &
      - \frac{N^2}{2g}[{\Delta_0}^2 - 2x(\Delta_0 - a_1g) + x^2(1-a_2g)]
      - N^2b_0ux \nonu
&   & - \frac{N^2}{2g'}[{\tau_0}^2 + 2u(\tau_0 - b_1g') + u^2(1-b_2g')]
      + F(x,u,N^2).
\ea
We keep the leading non-singular parts in $f(x,u)$.
This analytic part consists of the terms which are bilinear at most
with respect to $x$ and $u$.
These are necessary for applying the expansion around the saddle point
explained later.

We can regard the function $f(x,u)$ as an effective action for
a theory of baby universes.
The fields $x$ and $u$ effectively correspond to the operators
that make a puncture to a surface.
This puncture is characterized by the scaling operator.
Thus we distinguish the baby universe of the puncture operator from
the baby universe of the scaling operator ${\cal{O}}$.
The bilinear terms of $f(x,u)$ are related to the masses of these baby
universes.
We have two baby universes associated with the puncture operator $P$
and the scaling operator ${\cal{O}}$.
However the identification of these two baby universes is a little
complicated.
To make pure baby universes, it is convenient to rewrite $f(x,u)$
in the following form,
\ba
f(x,u) & = & N^2
         \left[
           \begin{array}{cc}
             \frac{\Delta_1}{g} & \frac{\tau_1}{g'}
           \end{array}
         \right]
         \left[
           \begin{array}{c}
             x \\ u
           \end{array}
         \right] - \frac{N^2}{2}
         \left[
           \begin{array}{c}
             x \\ u
           \end{array}
         \right]^{\rm T}
         \left[
           \begin{array}{cc}
             \frac{1}{g}-a_2 & b_0 \\
             b_0 & \frac{1}{g'}-b_2
           \end{array}
         \right]
         \left[
           \begin{array}{c}
             x \\ u
           \end{array}
         \right] \nonu
       &   & + F(x,u,N^2),
\label{BUEffAction}
\ea
where $\Delta_1 = \Delta_0-a_1g,\ \tau_1 = \tau_0-b_1g'$.
We drop irrelevant constant terms from the effective action.
The mass matrix of the baby universes
\be
M^2 = \left[
        \begin{array}{cc}
           \frac{1}{g}-a_2 & b_0 \\
           b_0 & \frac{1}{g'}-b_2
        \end{array}
      \right]
\ee
is diagonal only when $b_0 = 0$.
Then for $b_0\neq 0$, $x$ and $u$ are not eigenfunctions of the mass
matrix.
When $b_0 = 0$, the non-singular part has already diagonalized
and we can regard $x$ and $u$ as baby universe fields.
Whether $b_0 = 0$ or not depends on the choice of the representations
$\rho$ and $\rho'$ in (\ref{DefOfMMM}).
This parameter $b_0$ is nothing but the one which distinguishes
the generic case (Fig.\,1(a)) from the special case (Fig.\,1(b)).
We must discuss these two cases separately.

%
%
\subsection{$b_0\neq 0$}

For $b_0 \neq 0$ we diagonalize the non-singular part by redefining
$w = u + \frac{b_0g'}{1 - b_2g'}x$.
This redefinition is valid for $g' \neq {b_2}^{-1}$.
For a while we assume $g' \neq{b_2}^{-1}$ and consider the exceptional case
$g' = {b_2}^{-1}$ later.
Redefining the coupling constants as
\ba
\Delta & = & (\Delta_0 - a_1g) - \frac{b_0g}{1-b_2g'}(\tau_0 - b_1g'), \nonu
\tau   & = & \tau_0 - b_1g' ,
\ea
we find
\ba
f(x,u) & = & f(x,w) \nonu
& = & N^2
      \left[
        \begin{array}{cc}
          \frac{\Delta}{g} & \frac{\tau}{g'}
        \end{array}
      \right]
      \left[
        \begin{array}{c}
          x \\ w
        \end{array}
      \right] - \frac{N^2}{2}
      \left[
        \begin{array}{c}
          x \\ w
        \end{array}
      \right]^{\rm T}
      \left[
        \begin{array}{cc}
          \frac{1}{g}-a_2-\frac{{b_0}^2g'}{1-b_2g'} & 0 \\
          0 & \frac{1}{g'}-b_2
        \end{array}
      \right]
      \left[
        \begin{array}{c}
          x \\ w
        \end{array}
      \right] \nonu
&   & + F(x,w,N^2) .
\ea
The leading singular terms in $F(x,w,N^2)$ are the same
as the ones in $F(x,u,N^2)$.
The redefinition of the baby universe for the scaling operator ${\cal{O}}$
keeps the leading singular terms of the effective action.
When we seek for the double scaling limit of the modified model,
we must take care of the form of the singular part.
When we diagonalize the mass matrix of the baby universes,
the large $N$ expansion form of the singular part must be written as
(\ref{ScalingForm}) characterized by $\gamma$ and $d$.
This is the reason why we redefine only $u$.
The mass matrix $M^2$ can be diagonalized by redefining $x$.
However, this changes the form of the leading singular part.

As the non-singular part is diagonal in $x$ and $w$, we can read the
masses of these baby universes from each diagonal component,
\be
{m_w}^2 = \frac{1}{g'} - b_2 , \quad
{m_x}^2 = \frac{1}{g } - a_2 - \frac{{b_0}^2}{{m_w}^2} .
\ee
{}From assumption $w$ baby universe cannot be massless.
The last contribution to the mass square of $x$ baby universe comes from
the two point function of the puncture operator and the scaling operator
${\cal{O}}$.
We perform a large $N$ expansion around the saddle point $(x_s, w_s)$
which is given by the equation
$\frac{\del f}{\del x} = \frac{\del f}{\del w} = 0$.
The location of the saddle point is closely related to the masses of two baby
universes.
Thus we discuss separately by the sign of the mass eigenvalues


First we consider the case $g' > {b_2}^{-1}$.
This means that $w$ baby universe is tachyonic.
This case corresponds to the region above the dotted line in Fig.\,1(a).
It is easy to see there is no stable saddle point near $(x, w) = (0, 0)$.
The large $N$ expansion of the free energy which is found by expanding
$f(x,w)$ around the saddle point cannot be written
in the form of (\ref{ScalingForm}).
Thus this case causes scaling violation and we cannot take any double
scaling limit.
We can, however, take a continuum limit in which only planar graphs
survive.


Next we consider the case $g' < {b_2}^{-1}$ in which $w$ is massive.
The region below the dotted line in Fig.\,1(a) corresponds to this case.
We divide this case further into three: the $x$ baby universe is massive,
tachyonic and massless.


First the tachyonic case ${m_x}^2 < 0$ causes scaling violation for the same
reason as the tachyonic $w$ case.
In general when at least one of the baby universes is tachyonic,
we cannot take a double scaling limit.


The case in which both of the baby universe are massive corresponds
to the inside region of the hyperbolic curve in Fig.\,1(a).
The solution of the saddle point equation in this region is
\be
x_s = \frac{\Delta}{g{m_x}^2} + {\cal{O}}(\Delta^{1-\gamma}),\quad
w_s = \frac{\tau}{g'{m_w}^2} + {\cal{O}}(\Delta^{d+1-\gamma}).
\ee
Shifting the integration variables,
$\bar{x} = x - \frac{\Delta}{g{m_x}^2},\quad
 \bar{w} = w - \frac{\tau}{g'{m_w}^2}$
we have
\ba
\lefteqn{\log Z(\Delta, \tau, g, g')} \nonu
& = & \log\int_{-\infty}^\infty\A Nd\bar{x}
          \int_{-\infty}^\infty\A Nd\bar{w}
      \exp
      \left[
        -\frac{N^2{m_x}^2}{2}\bar{x}^2 -\frac{N^2{m_w}^2}{2}\bar{w}^2
      \right] \nonu
&   & \times\exp
      \left[
         F\left(
          \bar{x}+\frac{\Delta}{g{m_x}^2},
          \bar{w} + \frac{\tau}{g'{m_w}^2},N^2
          \right)
      \right] .
\ea
We rescale integration variables as
$\tilde{x} = \bar{x}N^{2/(2-\gamma)}{a_3}^{1/(2-\gamma)},\quad
 \tilde{w} = \bar{w}N^{2(1-d)/(2-\gamma)}$
and choose
$t = \frac{\Delta}{g{m_x}^2}N^{2/(2-\gamma)}{a_3}^{1/(2-\gamma)},\quad
 t_{\cal{O}} = \frac{\tau}{g'{m_w}^2}N^{2(1-d)/(2-\gamma)}$
as renormalized scaling variables.
Then we have
\ba
\lefteqn{\log Z(\Delta,\tau, g, g')} \nonu
& = & \log\int_{-\infty}^\infty\A N^{-\gamma/(2-\gamma)}d\tilde{x}\exp
      \left[
         \frac{1}{2}N^{-2\gamma/(2-\gamma)}{a_3}^{-2/(2-\gamma)}{m_x}^2
         \tilde{x}^2
      \right] \nonu
&   & \times\int_{-\infty}^\infty\A N^{(2d-\gamma)/(2-\gamma)}d\tilde{w}\exp
      \left[
         \frac{1}{2}N^{2(2d-\gamma)/(2-\gamma)}{m_w}^2\tilde{w}^2
      \right] \nonu
&   & \times \exp [F(\tilde{x} + t, \tilde{w} + t_{\cal{O}})] .
\ea
We keep $t$ and $t_{\cal{O}}$ finite and take the large $N$ limit.
Then the integrand becomes a product of delta-functions
$\delta(\tilde{x})\delta(\tilde{w})$.
We have finally
\be
\log Z(\Delta, \tau, g, g') = F(t, t_{\cal{O}}).
\ee
This scaling function $F(t, t_{\cal{O}})$ is the same as
(\ref{ScalingPP}) appeared in the $g = g' = 0$ theory.
Thus the modified theory belongs to the same universality class
of the unmodified theory.


In the massless case ${m_x}^2 = 0$, the location of the saddle point has
a new scaling nature,
\be
x_s = \left(\frac{(1-\gamma)\Delta}{ga_3}\right)^{\frac{1}{1-\gamma}}
      + {\cal{O}}(\Delta^{\frac{2+\gamma}{1-\gamma}}),\quad
w_s = \frac{\tau}{g'{m_w}^2} + {\cal{O}}(\Delta^{d+1-\gamma}).
\ee
We shift one of the integration variables,
$\bar{w} = w - \frac{\tau}{g'{m_w}^2}$.
The gaussian term of $\bar{w}$ becomes a delta-function $\delta(\bar{w})$
in the large $N$ limit.
Thus we can simply replace $w$ in the integrand by
$\frac{\tau}{g'{m_w}^2}$.
After this replacement we have
\be
\log Z(\Delta, \tau, g, g') = \log\int_{-\infty}^\infty\A Ndx\e^{f(x)},
\label{EffTheoryOfX}
\ee
\be
f(x) = \frac{N^2\Delta}{g}x + F\left(x, \frac{\tau}{g'{m_w}^2},N^2\right) .
\ee
The problem reduces to a theory of a single baby universe.
We expand the effective action $f(x)$ around the saddle point $x_s$
and find the $N^2$-order contribution
\ba
\log Z(\Delta, \tau, g, g')
& = & N^2\left(
           \frac{1}{2-\gamma}{a_3}^{-\frac{1}{1-\gamma}}
           \left(
              \frac{(1-\gamma)\Delta}{ga_3}
           \right)^{\frac{2-\gamma}{1-\gamma}}
           + \cdots
         \right)\nonu
&   & + N^2\frac{\tau}{g'{m_w}^2}
         \left(
           -b_3
           \left(
              \frac{(1-\gamma)\Delta}{ga_3}
           \right)^{\frac{d+1-\gamma}{1-\gamma}}
           + \cdots
         \right) \nonu
&   & + \frac{1}{2}N^2\left(\frac{\tau}{g'{m_w}^2}\right)^2
         \left(
           -b_4
           \left(
              \frac{(1-\gamma)\Delta}{ga_3}
           \right)^{\frac{2d-\gamma}{1-\gamma}}
           + \cdots
         \right) \nonu
&    & + \cdots + {\cal{O}}(N^0) .
\ea
This form of the large $N$ expansion is similar to (\ref{ScalingForm})
if we take new string susceptibility $\bar{\gamma}$ and
new gravitational dimension $\bar{d}$ of the scaling operator as
\be
\bar{\gamma} = \frac{\gamma}{\gamma-1},\quad
\bar{d}      = \frac{\gamma-d}{\gamma-1} .
\ee
These new values can be understood as the result of
the change of the branch in $\alpha$ in the Liouville interaction
$\e^{\alpha\phi}$ from $\alpha_+$ to $\alpha_-$.
For the string susceptibility, we explained it in \S\ref{Intro}.
For the gravitational dimension, we can easily check
$\bar{d} = 1 - \frac{\beta_+}{\alpha_-}$ agrees with the above result.
The branch for $\beta$ is unchanged.

The massless $x$ region lies on the hyperbolic curve in Fig.\,1(a).
We consider this curve as a critical line.
When we fix the value of $g'$, the location of the critical point $g_t$
varies with $g'$.
Of course, if we turn off the square term of the scaling operator
${\cal{O}}$, $g_t = {a_2}^{-1}$.

It is easy to see that the double scaling limit is realized by choosing new
scaling variables as
\ba
\bar{t}      & = & \frac{\Delta}{g} N^{2/(2-\bar{\gamma})}
\left((1-\bar{\gamma})a_3\right)^{-(1-\bar{\gamma})/(2-\bar{\gamma})}, \nonu
t_{\cal{O}}  & = & \frac{\tau}{g'{m_w}^2} N^{2(1-\bar{d})/(2-\bar{\gamma})}
               =   \frac{\tau}{g'{m_w}^2} N^{2(1-d)/(2-\gamma)} .
\ea
In the double scaling limit the free energy acquires new scaling nature
\be
\log Z(\Delta,\tau,g,g') = \bar{F}(\bar{t}, t_{\cal{O}}),
\ee
\be
\bar{F}(\bar{t}, t_{\cal{O}})
= \frac{\bar{t}^{2-\bar{\gamma}}}{2-\bar{\gamma}}
  - \bar{a}_1\log\bar{t} - \sum_{h=2}^\infty\bar{a}_h
  \bar{t}^{(2-\bar{\gamma})(1-h)}
  - \sum_{h=0}^\infty\sum_{k=1}^\infty\bar{b}_{h,k}
  \bar{t}^{(2-\bar{\gamma})(1-h)}
  \left(t_{\cal{O}}\bar{t}^{\bar{d}-1}\right)^k .
\label{ScalingMP}
\ee
We can derive a relation between the modified free energy $\bar{F}$
and the unmodified one $F$ appeared in (\ref{ScalingPP}).
To do this we rescale the integration variable in (\ref{EffTheoryOfX})
as $t = x N^{2/(2-\gamma)}{a_3}^{1/(2-\gamma)}$ and take the double
scaling limit.
We find a formula similar to the one derived in \cite{KLEHASHI}
\be
\bar{F}(\bar{t}, t_{\cal{O}}) = \log\int_{-\infty}^\infty\A dt\,
\e^{\epsilon t\bar{t} + F(t,t_{\cal{O}})},
\label{RelMP}
\ee
where $\epsilon$ is a constant which depends on only $\bar{\gamma}$
(or $\gamma$),
$\epsilon = (1-\bar{\gamma})^{(1-\bar{\gamma})/(2-\bar{\gamma})}$ .
This relation can be considered as a nonperturbative definition of
the new scaling function $\bar{F}$.


Now we come back to the case $g' = {b_2}^{-1}$ we have left at the
beginning of this subsection.
The baby universe $w$ is not well-defined for this case.
As there is no quadratic term of $u$ in the effective action $f(x,u)$,
we cannot absorb the term $-N^2b_0xu$ by redefining the baby universe $u$.
For finite $g$, the result of the large $N$ expansion around the saddle
point cannot be written in the form of (\ref{ScalingForm}) by any means.
This means that we cannot take a double scaling limit in this case.
We cannot regard the modified theory in this parameter region
as an effective theory of baby universes.
When $g = 0$ which means the baby universe corresponding to the puncture
operator is absent or extremely heavy, we can find a new scaling nature
in the theory.
To show this we return to the starting point,
\be
\log Z(\Delta_0, \tau_0, 0, b_2^{-1})
= \log N\int_{-\infty}^\infty\A du\,\e^{f(u)},
\ee
\be
f(u) = N^2b_2u\left(\tau_0 - \frac{b_1}{b_2} - \frac{b_0}{b_2}\Delta_0\right)
       + F(\Delta_0,u,N^2) .
\ee
As we have no quadratic terms of $u$, the $u$ baby universe is massless.
Defining new coupling constants as
$\Delta = \Delta_0,\quad
\tau = \tau_0-\frac{b_1}{b_2}-\frac{b_0}{b_2}\Delta_0$,
we find the location of the saddle point
\be
u_s = -\frac{b_2}{b_4}\tau\Delta^{-2d+\gamma} + {\cal{O}}(\Delta^{1-d}).
\ee
This estimate of the saddle point is valid for sufficiently small $\Delta$.
Integrating around $u_s$ and choosing the scaling variables as
\be
t = \Delta N^{2/(2-\gamma)}{a_3}^{1/(2-\gamma)},\quad
\bar{t}_{\cal{O}} = b_2\tau N^{2(1-\bar{d})/(2-\gamma)},\quad
\bar{d} = \gamma - d,
\ee
we find the modified free energy in the double scaling limit
\be
\log Z(\Delta, \tau, 0, b_2^{-1}) = \bar{F}(t,\bar{t}_{\cal{O}}) ,
\ee
\ba
\bar{F}(t,\bar{t}_{\cal{O}}) & = & -\frac{t^{2-\gamma}}{(2-\gamma)(1-\gamma)}
    -\bar{a}_1\log t - \sum_{h=2}^\infty\bar{a}_h t^{(2-\gamma)(1-h)} \nonu
&   & -\sum_{h=0}^\infty\sum_{k=1}^\infty\bar{b}_{h,k}t^{(2-\gamma)(1-h)}
      \left(\bar{t}_{\cal{O}}t^{\bar{d}-1}\right)^k .
\label{ScalingPM}
\ea
Thus the string susceptibility of the modified theory is the same as
the one before modification.
On the other hand, the scaling operator ${\cal{O}}$ acquires a new
gravitational dimension $\bar{d} = \gamma - d$.
In language of Liouville theory, this is explained by the change
of the branch in the Liouville dressing $\e^{\beta\phi}$, that is,
\be
\bar{d} = \gamma - d = 1 - \frac{\beta_-}{\alpha_+} .
\ee
This universality appears at a single point $(g, g') = (0, {b_2}^{-1})$
in Fig 1(a).
In a sense we can say the value $g' = {b_2}^{-1}$ is special.
Thus we may call this special value of $g'$ as a critical point for $g'$.

Rescaling the integration variable as $t_{\cal{O}} = uN^{2(1-d)/(2-\gamma)}$,
we find that the modified free energy is related to the unmodified one
\be
\bar{F}(t,\bar{t}_{\cal{O}}) = \log\int_{-\infty}^\infty\A dt_{\cal{O}}\,
    \e^{t_{\cal{O}}\bar{t}_{\cal{O}} + F(t,t_{\cal{O}})}.
\ee
This relation resembles (\ref{RelMP}).
Now the integration is performed over $t_{\cal{O}}$ not $t$.

So far we have studied the generic case $b_0 \neq 0$.
It is worth noting that for $b_0 \neq 0$, the change of the branch in
$\alpha$ and $\beta$ never occurs at the same time.
This is deeply related to the fact that we cannot make the two
baby universes corresponding to the puncture operator and the scaling
operator massless at the same time.

%
%
\subsection{$b_0 = 0$}

Now we study the special case $b_0 = 0$ when the mass matrix of baby
universes is diagonal.
The universality diagram for this special case is shown in Fig.\,1(b).
A remarkable difference from the generic case is that it is possible to
fine tune $g$ and $g'$ simultaneously.
This means that the modified theory may have another new scaling nature
which is present in the generic case.
The simultaneous tuning corresponds to the single point
$(g,g')=({a_2}^{-1},{b_2}^{-1})$ in Fig.\,1(b).
The work of Klebanov and Hashimoto \cite{KLEHASHI} covers this special case.
The non-singular part of the effective action
$f(x,u)$ of baby universes (\ref{BUEffAction}) has already diagonalized.
We can treat $x$ and $u$ as pure baby universes.
We interpret that $x$ corresponds to the puncture operator and $u$ to
the scaling operator ${\cal{O}}$.
The mass squares of these baby universes are read from the effective action,
\be
{m_x}^2 = \frac{1}{g}  - a_2 ,\qquad
{m_u}^2 = \frac{1}{g'} - b_2 .
\ee
As in the generic case, we divide into five cases by the sign of
the mass squares.
In most cases the analysis is common to the generic one and
the answer is the same as \cite{KLEHASHI}.
We do not repeat in detail and show only some essential results.


First we consider the case where $x$ and $u$ are massive.
This case corresponds to the inside region of the rectangle in Fig.\,1(b).
In this case the integral becomes a product of delta-functions around
the saddle point in the large $N$ limit.
This allows one to replace $x$ with $x_s = \frac{\Delta}{1-a_2g}$
and $u$ with $u_s = \frac{\tau}{1-b_2g'}$,
where $\Delta = \Delta_0 - a_1g$, $\tau = \tau_0 - b_1g'$.
Therefore the modified theory is in the same universality class
as the unmodified one.
We select the renormalized scaling variables,
\be
t           = \frac{\Delta}{1-a_2g}N^{2/(2-\gamma)}{a_3}^{1/(2-\gamma)},\quad
t_{\cal{O}} = \frac{\tau}{1-b_2g'} N^{2(1-d)/(2-\gamma)} .
\ee
In the double scaling limit, the sum over surfaces converges to
$F(t, t_{\cal{O}})$, the generating function of the $g = g' = 0$ theory.


The second case is $g = {a_2}^{-1},\ 0 \leq g' < {b_2}^{-1}$, \ie,
the case of massless $x$ and massive $u$.
Contrary to the generic case, the massless condition ${m_x}^2 = 0$
requires that the coupling constant $g$ be fine tuned to ${a_2}^{-1}$.
The double scaling limit is realized by choosing scaling variables as
\be
\Delta = \Delta_0 - \frac{a_1}{a_2},\quad
\tau   = \tau_0   - b_1g' ,
\ee
\be
\bar{t} = a_2\Delta^{2/(2-\bar{\gamma})}
          {(1-\bar{\gamma})a_3}^{-(1-\bar{\gamma})/(2-\bar{\gamma})},\quad
t_{\cal{O}} = \frac{\tau}{1-b_2g'}N^{2(1-\bar{d})/(2-\bar{\gamma})} ,
\ee
where $\bar{\gamma}$ and $\bar{d}$ are the new string susceptibility and
the new gravitational dimension of the modified theory,
\be
\bar{\gamma} = \frac{\gamma}{\gamma-1},\quad
\bar{d}      = \frac{\gamma-d}{\gamma-1} .
\ee
These values are explained by the change of branch in $\alpha$.
The massless condition for $x$ baby universe is nothing but the fine tuning
of the coupling constant $g$ to its critical value $g_t = {a_2}^{-1}$.
As the consequence of this fine tuning, the branch in $\alpha$ changes.
In the double scaling limit, the free energy becomes
$\bar{F}(\bar{t},t_{\cal{O}})$ which appeared in (\ref{ScalingMP})
in the generic case.
Thus the modified theory belongs to the same universality class as the case
${m_x}^2 = 0$ and ${m_w}^2 > 0$ in the generic case.


The third case is $0 \leq g < {a_2}^{-1}$ and $g' = {b_2}^{-1}$,\ie,
the case of massive $x$ and massless $u$.
When $b_0 \neq 0$, the coupling constant $g$ must be zero to take a
double scaling limit.
In this case, however, $g$ is not restricted to zero.
The scaling variables are defined by
\be
\Delta = \Delta_0 - a_1g, \quad
\tau   = \tau_0 - \frac{b_1}{b_2},
\ee
\be
t = \frac{\Delta}{1-a_2g}N^{2/(2-\gamma)}{a_3}^{1/(2-\gamma)}, \quad
\bar{t}_{\cal{O}} = b_2\tau N^{2(1-\bar{d})/(2-\gamma)} .
\ee
The scaling operator ${\cal{O}}$ gets a new gravitational dimension
$\bar{d} = \gamma - d$.
This new gravitational dimension is understood by the change of the branch
in $\beta$.
The scaling function $\bar{F}(t,{\cal{O}})$ is the same as (\ref{ScalingPM})
in the generic case.


The fourth case corresponds to simultaneous tuning of the two coupling
constants $g$ and $g'$.
The solution of the saddle point equation is
\be
x_s = \left(\frac{(1-\gamma)a_2\Delta}{a_3}\right)^{\frac{1}{1-\gamma}}
      + {\cal{O}}(\Delta^{\frac{2+\gamma}{1-\gamma}}), \quad
u_s = -\frac{b_2}{b_4}\tau\Delta^{-2d+\gamma}
      + {\cal{O}}(\Delta^{1-d}),
\ee
where $\Delta$ and $\tau$ are given by
\be
\Delta = \Delta_0 - \frac{a_1}{a_2}, \quad
\tau   = \tau_0 - \frac{b_1}{b_2}.
\ee
If we denote the derivative of the function $f(x,u)$ with $x$ by
$f'(x,u)$ and the derivative with $u$ as $\dot{f}(x,u)$, we have
\be
\log Z(\Delta,\tau,{a_2}^{-1},{b_2}^{-1})
= \left[
   f - \frac{1}{2}\log\left(\frac{f''\ddot{f}-(\dot{f}')^2}{N^4}\right)
  \right]_{x = x_s, u = u_s} + {\cal{O}}(N^{-2}) .
\ee
After some calculation we find new scaling nature in the free energy.
The double scaling limit is realized by selecting the scaling variables as
\be
\bar{t} = a_2\Delta^{2/(2-\bar{\gamma})}
         {(1-\bar{\gamma})a_3}^{-(1-\bar{\gamma})/(2-\bar{\gamma})}, \quad
\bar{t}_{\cal{O}} = b_2\tau N^{2(1-\bar{d})/(2-\bar{\gamma})}.
\ee
The new string susceptibility $\bar{\gamma}$ and
the new gravitational dimension $\bar{d}$ are
\ba
\bar{\gamma} & = & \frac{\gamma}{\gamma - 1} = 2 + \frac{Q}{\alpha_-},\nonu
\bar{d}      & = & \frac{d}{\gamma - 1} = 1 - \frac{\beta_-}{\alpha_-}.
\ea
These new values indeed correspond to the simultaneous change of
the branches in $\alpha$ and $\beta$.
Keeping these two scaling variables finite and letting $N$ to infinity,
we find the scaling function of the form,
\be
\log Z(\Delta, \tau,{a_2}^{-1}, {b_2}^{-1})
= \bar{F}(\bar{t}, \bar{t}_{\cal{O}}),
\ee
\be
\bar{F}(\bar{t}, \bar{t}_{\cal{O}})
= \frac{t^{2-\bar{\gamma}}}{2-\bar{\gamma}} - \bar{a}_1\log\bar{t}
  - \sum_{h=2}^\infty\bar{a}_h\bar{t}^{(2-\bar{\gamma})(1-h)}
  - \sum_{h=0}^\infty\sum_{k=1}^\infty\bar{b}_{h,k}
  \bar{t}^{(2-\bar{\gamma})(1-h)}
  \left(\bar{t}_{\cal{O}}\bar{t}^{\bar{d}-1}\right)^k .
\label{ScalingMM}
\ee
The scaling function (\ref{ScalingMM}) may be expressed by
an integral over scaled coupling constants,
\be
t = xN^{2/(2-\gamma)}{a_3}^{1/(2-\gamma)}, \quad
t_{\cal{O}} = uN^{2(1-d)/(2-\gamma)} .
\ee
We find a simple relation
\be
\bar{F}(\bar{t}, \bar{t}_{\cal{O}}) =
\log\int_{-\infty}^\infty\A dt\int_{-\infty}^\infty\A dt_{\cal{O}}\,
\e^{\epsilon t\bar{t} + t_{\cal{O}}\bar{t}_{\cal{O}} + F(t, t_{\cal{O}})},
\ee
where $\epsilon = (1-\bar{\gamma})^{(1-\bar{\gamma})/(2-\bar{\gamma})}$
is a constant which depends on the universality of the unmodified theory.
The appearance of this new universality class is a feature of
the $b_0 = 0$ case.


The last case is the one where at least one of the baby universes
is tachyonic.
This corresponds to the outside region of the rectangle in the
universality diagram Fig.\,1(b).
As there are no stable saddle points near $x = u = 0$, we cannot take any
double scaling limit.
Thus we are not interested in this case.

\subsection{Scaling operators with $d > \frac{1}{3}(1+\gamma)$}

Finally, we briefly comment on the case
$d > \frac{1}{3}(1+\gamma)$.
In this case, the effective action of baby universes cannot be written in
the form (\ref{BUEffAction}).
There are some extra higher-order terms in the non-singular part.
Nevertheless when $(1-a_2g)(1-b_2g')-{b_0}^2gg' > 0$, we can take a
double scaling limit by choosing the scaling variables as
\ba
t & \propto & \Delta N^{2/(2-\gamma)},\quad
\Delta = \Delta_1 - \frac{b_0g}{1-b_2g'}\tau_1, \nonu
t_{\cal{O}} & \propto & \tau N^{2(1-d)/(2-\gamma)},\quad
\tau = \tau_1 - \frac{b_0g'}{1-a_2g}\Delta_1,
\ea
where $\Delta_1$ and $\tau_1$ are appeared in (\ref{BUEffAction}).
In the double scaling limit, the modified theory is in the same universality
class as the unmodified theory.
Such a double scaling limit is also possible when
$d \leq \frac{1}{3}(1+\gamma)$.
The condition $(1-a_2g)(1-b_2g')-{b_0}^2gg' > 0$ corresponds to the case
both $x$ and $w$ baby universes are massive.
Therefore we have two different ways of taking the double scaling limit
when $(1-a_2g)(1-b_2g')-{b_0}^2gg' > 0$ and $d \leq \frac{1}{3}(1+\gamma)$.
We cannot take a double scaling limit out of this region unless we turn
off the coupling constant of scaling operators with
$d > \frac{1}{3}(1+\gamma)$.

The reason why we are not interested in the scaling operators whose
gravitational dimension satisfies $d > \frac{1}{3}(1+\gamma)$
is that we cannot change the branch in $\alpha$ or $\beta$ in a
double scaling limit.
The square term of such operator is irrelevant when the baby universe
associated with the operator is massive.
Otherwise it destroys scaling property of the theory
in the continuum limit.

%
\section{Representation of Scaling Operators and the Modified Matrix Models}

In appendix A we have studied the 1-matrix model with
potential of sixth order.
We have learned that both $\tr\Phi^4$ and
$\tr(\Phi^6+\alpha\Phi^4),\ \alpha\neq -15$ behave as the puncture
operator in the continuum limit.
This example teaches us that the way to represent scaling operators by
matrices is not unique.
For a while we discuss the problems of the representation of the scaling
operators.

In the 1-matrix model of appendix A, a special linear
combination $\tr(\Phi^6-15\Phi^4)$
behaves as the scaling operator with dimension $\frac{1}{3}$.
In general, we can produce an unknown matrix representation of a scaling
operator like this way.
{}From two representations $\rho_1, \rho_2$ of a scaling operator with
gravitational dimension $d$, we can make a new representation
$\rho = \rho_1 + r\rho_2$ so that it represents a scaling operator
with dimension larger than $d$ by choosing $r$ appropriately.
It is essential that the new representation is always a linear combination
of the two representations.
This fact comes from the universality of the matrix models, that is,
the form of the scaling function is unchanged up to the dilatation of the
scaling variables.
Thus we can cancel the leading singular terms in the one point function
of these scaling operators
$\bigl<\rho_i\bigr>,\ i=1,2$ by defining a linear combination
$\rho = \rho_1 + r\rho_2$.

In the last section, we have learned that the scaling nature of the modified
matrix model depends on not only $g$ and $g'$ but also $b_0$.
The value of $b_0$ is determined by the choice of the representations
$\rho(P)$ and $\rho'({\cal{O}})$.
Is it be possible to choose these representations so that $b_0$ vanishes?
The answer of this question is yes.
The two point function $\bigl<\rho(P)\rho'({\cal{O}})\bigr>$ of the
unmodified model determines the value of $b_0$.
Even if $b_0\neq 0$, we can find a new representation of the scaling operator
${\cal{O}}$ that makes the mass matrix of the baby universes diagonal.
To do this, we need another representation of the scaling operator
$\rho''({\cal{O}})$ such that the two point function
$\bigl<\rho(P)\rho''({\cal{O}})\bigr>$ does not vanish.
Such a representation always exists.
We define a new representation of the scaling operator by a linear
combination of these, $\bar{\rho}=\rho'+r\rho''$.
If we replace $\rho'$ by this $\bar{\rho}$, the parameter $b_0$
must be a linear function of $r$.
Thus we can make $b_0=0$ by appropriately selecting $r$.
This means that we can always fine tune $g$ and $g'$ simultaneously
for the scaling operator whose gravitational dimension satisfies
the condition $d \leq \frac{1}{3}(1+\gamma)$.

%
\section{Discussion}

In this paper we have confirmed that the string susceptibility and
the gravitational dimension in the double scaling limit are understood
by the idea that a fine tuning of square terms changes the branch of
Liouville dressing of some scaling operators.
This fine tuning is possible when the mass matrix of the baby
universes is already diagonal.
Even if it is not diagonal, we can diagonalize it by redefining the baby
universe corresponding to a scaling operator except the puncture operator.
When the mass matrix is diagonalized by this procedure, the
simultaneous tuning of square terms is impossible.
Our new discovery is that the representations $\rho(P)$ and
$\rho'({\cal{O}})$ have a effect not only in the critical values for
$g$ and $g'$ but also in the possibility of simultaneous change of the
branch in $\alpha$ and $\beta$.

The existence of an upper bound in the gravitational dimension of
the scaling operator above which we cannot fine tune the coupling constants
of square terms is interesting for us.
This upper bound means that the branch of
the Liouville dressing for such operators cannot be changed in the
modified theories we have studied.
If we fix the type of the minimal conformal matter coupled to gravity,
the number of allowed operators is always smaller than the number of
the gravitational primary fields.
We think this is related to the fact that the difference between
the behavior of $\e^{\beta_+\phi}$ and $\e^{\beta_-\phi}$ becomes
large for such operators in the weakly interacting region of large
$\phi$ in Liouville theory.
Our result doesn't insist that it is impossible to change the branch of
such operators by any means.
It may be realized by another type of modified matrix models.

We have restricted our argument to the case where there are at most
two square terms, \ie, $m=1$ in (\ref{ModifiedMM}).
Of course we can generalize to the case $m\geq 2$.
Then the size of the mass matrix of the baby universes becomes $(m+1)$.
The simultaneous tuning of $(m+1)$ square terms is possible only when
the mass matrix is diagonal and all the scaling operators satisfy
the condition $d \leq \frac{1}{3}(1+\gamma)$.
Choosing the representations of scaling operators appropriately,
we can always make the mass matrix diagonal.
The idea that a fine tuning changes the branch of Liouville dressing
of some scaling operators is valid as well.
It is interesting that the matrix models we have considered and the theory
of baby universes are intimately related like this.

When the square terms are fine tuned, the continuum theory doesn't have
semiclassical limit.
This fact makes the modified matrix models more complex.
However, the results of the modified theories strongly suggest
the existence of such a strange Liouville theory.
We hope that physical aspect of these fine tuned theories becomes clear.
Our work will be easily applied to the $c=1$ matrix model.

%
%
\par
\begin{flushleft}
\bf Acknowledgement\\[2mm]
\end{flushleft}
We are grateful to H. Itoyama for reading the manuscript.
This work is supported in part by Grant-in-Aid for Scientific Research from
the Ministry of Education, Science and Culture.

\setcounter{section}{0}
\setcounter{equation}{0}
%
%
\appendix{General Form of the Results of the Matrix Integration}

To study the modified matrix models, it is necessary to know a general
form of the free energy expanded around a given critical point.
In this appendix we study a pedagogical example and deduce the general
form of the free energy from this example.
We use the 1-matrix model with potential of the sixth order as the example.
This is the simplest model if we want to include a source term of
a scaling operator with gravitational dimension greater than $0$.
The partition function is defined by the integration over an $N \times N$
hermitian matrix $\Phi$,
\be
Z(\lambda, \kappa) = \int\A D\Phi\,\e^{-NS\tr(\half\Phi^2 - \lambda\Phi^4
- \kappa\Phi^6)},
\ee
where $\lambda$ and $\kappa$ are parameters.
Usually we relate $\lambda$ with the cosmological constant in the Liouville
theory.
It is well-known that the large-$N$ expansion of the free energy corresponds
to the topological expansion,
\be
\log \frac{Z(\lambda, \kappa)}{Z(0,0)} = \sum_{h=0}^\infty N^{2-2h}
\,e^{(h)}(\lambda, \kappa).
\ee

Using the method of orthogonal polynomials or the steepest descent method,
we can find the contribution of planar graphs,
\ba
e^{(0)}(\lambda, \kappa) & = & - 36\lambda\kappa a^{10}
                               + (15\kappa - 6 \lambda^2)a^8
                               + (3\lambda + 20\kappa)a^6
                               + \left(5\lambda - \frac{1}{6}\right)a^4 \nonu
                         &   & - \frac{2}{3}a^2 + \frac{5}{6}
                               + \frac{1}{2}\log a^2 ,
\label{Planar}
\ea
where $a$ satisfies an equation
\be
60\kappa a^6 + 12\lambda a^4 - a^2 + 1 = 0 .
\label{NormalizationCond}
\ee
{}From (\ref{Planar}) and (\ref{NormalizationCond}) we eliminate $\lambda$
and find
\be
e^{(0)}(\lambda, \kappa)
= \frac{1}{24}(1 - a^2)(9 - a^2) + \frac{1}{2}\log a^2
  - 7\kappa a^6 + 2\kappa a^8 + 30 \kappa^2 a^{12} .
\ee
This free energy shows a singular behavior if we fine tune
$\lambda$ and $\kappa$.
The critical point such that (\ref{NormalizationCond}) has triple root
in $a^2$ is known as the $k=3$ critical point of 1-matrix model.
At this critical point, the theory is identified with the Liouville theory
coupled with $(2, 5)$ minimal conformal matter.
The $k=3$ critical point is achieved by fine tuning the parameters
to the values,
\[
\kappa_c = -\frac{1}{1620},\quad\lambda_c = \frac{1}{36},\quad a_c^2=3 .
\]
To find the singular behavior for the free energy near the critical
point, it is convenient to introduce
$a^2 = a_c^2(1 - b),\ \Delta = \lambda_c - \lambda$.
$\Delta$ is nothing but the cosmological constant.
We need the expansion of $b$ by $\Delta^{\frac{1}{3}}$ up to the seventh
order,
\[
b =  3(4\Delta)^{\frac{1}{3}} - 6(4\Delta)^{\frac{2}{3}}
   + 9(4\Delta) - 10(4\Delta)^{\frac{4}{3}} + 7(4\Delta)^{\frac{5}{3}}
   - \frac{22}{3}(4\Delta)^{\frac{7}{3}} + {\cal{O}}(\Delta^{\frac{8}{3}}) .
\]
Defining $\tau = \kappa_c - \kappa$, we expand $e^{(0)}(\Delta, \tau)$
around $\Delta = \tau = 0$.
The $\tau$-independent contribution is
\ba
e^{(0)}(\Delta, 0)
&\! = \! &  \left(-\frac{19}{40} + \frac{1}{2}\log{3}\right) - \frac{b^3}{20}
      - \frac{b^4}{10} - \frac{3b^5}{20} - \frac{3b^6}{40}
      - \frac{b^7}{14} + {\cal{O}}(b^8) \nonu
&\! = \! &  \left(-\frac{19}{40} + \frac{1}{2}\log{3}\right)
      - \frac{27}{5}\Delta
      + 1458\Delta^2 - \frac{19683}{28}(4\Delta)^{\frac{7}{3}} \nonu
&        & + {\cal{O}}(\Delta^{\frac{8}{3}}).
\ea
As the leading singular term behaves as $\Delta^{2-\gamma}$,
the string susceptibility is $-\frac{1}{3}$.
This exponent suggests that the continuum limit of this matrix model
is identified with the Liouville theory coupled to $(2, 5)$ minimal
conformal matter.

The singular behavior for the linear and quadratic terms in $\tau$
is derived by differentiating with respect to $\tau$
\be
\left. \frac{\del e^{(0)}}{\del\tau} \right|_{\tau = 0}
= - 27 + 34992\Delta - 98415(4\Delta)^{\frac{4}{3}}
  + {\cal{O}}(\Delta^{\frac{5}{3}}),
\ee
\be
\left. \frac{\del^2 e^{(0)}}{\del\tau^2} \right|_{\tau = 0}
= 437400 - 7873200(4\Delta)^{\frac{1}{3}} + {\cal{O}}(\Delta^{\frac{2}{3}}) .
\ee
Gathering these contributions, we arrive at the final form of the free energy,
\ba
e^{(0)}(\Delta, \tau)
& = & - \frac{27}{5}\Delta + \frac{1}{2}\cdot 2196\Delta^2 - 27\tau
      + \frac{1}{2}\cdot 437400\tau^2 + 34992\Delta\tau \nonu
&   & - \frac{19683}{28}(4\Delta)^{\frac{7}{3}} + \cdots
      - 98415(4\Delta)^{\frac{4}{3}}\tau + \cdots \nonu
&   & - \frac{1}{2}\cdot 7873200(4\Delta)^{\frac{1}{3}}\tau^2 + \cdots \nonu
&   & + {\cal{O}}(\Delta^{-\frac{2}{3}}\tau^3) .
\ea
{}From the definition of the gravitational scaling dimension,
one can easily find that $\tr \Phi^6$ corresponds to the
puncture operator $(d=0)$ in the continuum limit.

It is well-known that in the $(2, 5)$ minimal matter theory there are two
primaries: the identity operator and a operator with conformal dimension
$-\frac{1}{5}$.
After coupling with gravity, the latter becomes the puncture operator
and the dressed identity operator gain the gravitational dimension
$\frac{1}{3}$.
We want to replace $\tr\Phi^6$ by the dressed identity operator.
The operator with $d = \frac{1}{3}$ should be represented by a linear
combination of these operators.
Thus we replace $\tr\Phi^6$ by a trial operator
$\tr{\cal{O}}_\alpha = \tr(\Phi^6+\alpha\Phi^4)$ and determine $\alpha$ so
that ${\cal{O}}_\alpha$ behaves as the scaling operator with
$d = \frac{1}{3}$.
To this end we consider a matrix integral,
\be
Z(\Delta, \tau) = \int\A D\Phi\,
\e^{-N\tr(\half\Phi^2-\lambda_c\Phi^4-\kappa_c\Phi^6+\Delta\Phi^4
    +\tau{\cal{O}}_\alpha)} .
\label{GeneratingFunc}
\ee
It is not necessary to repeat calculation.
What we have to do is just replacing
$\lambda$ with $\lambda + \alpha(\kappa - \kappa_c)$ in the previous result.
As the location of the critical point doesn't move, the definition of the
scaling parameters $\Delta$ and that of $\tau$ are the same as before.
After some calculations we find the expansion form of $e^{0}(\Delta, \tau)$,
\ba
\lefteqn{e^{(0)}(\Delta,\tau)} \nonu
& = & - \frac{27}{5}\Delta + \frac{1}{2}\cdot 2196\Delta^2
      - 27\left(\frac{\alpha}{5}+1\right)\tau
      + \frac{1}{2}\cdot 2196(\alpha^2+24\alpha+150)\tau^2 \nonu
&   & + 2196(\alpha+12)\Delta\tau
      - \frac{9}{28}(108)^{\frac{7}{3}}\Delta^{\frac{7}{3}} + \cdots \nonu
&   & + \tau\left[
-\frac{4}{3}(\alpha+15)(108)^{\frac{7}{3}}\Delta^{\frac{4}{3}}
               +
\frac{1}{2}\left(\frac{13}{5}\alpha+48\right)(108)^{\frac{8}{3}}
               \Delta^{\frac{5}{3}} + \cdots
            \right] \nonu
&   & + \frac{1}{2}\tau^2
        \left[-(\alpha+15)^2(108)^{\frac{7}{3}}\Delta^{\frac{1}{3}}
        + \frac{1}{6}(\alpha+15)(13\alpha+285)(108)^{\frac{8}{3}}
        \Delta^{\frac{2}{3}}\right. \nonu
&   & \left. - \frac{5}{3}(108)^3(2\alpha^2+90\alpha+945)\Delta+\cdots\right]
      \nonu
&   & + {\cal{O}}(\Delta^{-\frac{2}{3}}\tau^3) .
\label{ScalingUpto2ndOrder}
\ea
If we set $\alpha = -15$, the terms proportional to $\Delta^{\frac{4}{3}}\tau$,
$\Delta^{\frac{1}{3}}\tau^2$ and $\Delta^{\frac{2}{3}}\tau^2$ vanish
and the operator $\tr{\cal{O}}_{-15} = \tr(\Phi^6 - 15\Phi^4)$
behaves as the operator desired.
It is worth noting that this $e^{(0)}(\Delta, \tau)$ contains odd-order terms
in $\tau$ which is absent in the paper \cite{KLEHASHI}.
They assumed that odd-point function of the scaling operator vanish.
Our example shows that this is not generally the case.

Our result (\ref{ScalingUpto2ndOrder}) offers a general form for
the generating function of the correlation function of a scaling operator
${\cal{O}}$.
We consider $\Delta$ as the cosmological constant and $\tau$ as a coupling
constant of the scaling operator and define the matrix integral
$Z(\Delta, \tau)$ as (\ref{GeneratingFunc}).
We fine tune parameters of the matrix model so that the string susceptibility
becomes $\gamma$.
We choose the matrix representation of ${\cal{O}}$ so that
the gravitational dimension becomes $d$ at this criticality.
In general the free energy can be written in the following form
\[
\log Z(\Delta, \tau) = \mbox{Analytic Part}\,+ F(\Delta,\tau,N^2),
\]
\ba
F(\Delta,\tau,N^2)
& = & N^2\left(-\frac{a_3}{(2-\gamma)(1-\gamma)}
      \Delta^{2-\gamma} + \cdots\right)
      - N^0(a_4\log\Delta + \cdots) \nonu
&   & - N^{-2}(a_5\Delta^{-2+\gamma} + \cdots) + {\cal{O}}(N^{-4}),\nonu
&   & - N^2\tau(b_3\Delta^{d+1-\gamma} + \cdots)
      - \half N^2\tau^2(b_4\Delta^{2d-\gamma}
        + \cdots) \nonu
&   & - {\textstyle{1\over 6}}N^2\tau^3(b_5\Delta^{3d-1-\gamma} + \cdots)
      + \cdots + {\cal{O}}(N^0\tau\Delta^{d-1}) .
\ea
Here we separate non-singular part from the singular part
$F(\Delta,\tau,N^2)$.
The structure of the singular part is indeed consistent with the result
our example (\ref{ScalingUpto2ndOrder}).
This scaling property of the singular part comes from the fact
that one can take a double scaling limit.
The double scaling limit is achieved by choosing the renormalized
scaling variables as
\be
t = \Delta N^{2/(2-\gamma)}{a_3}^{1/(2-\gamma)},\
t_{\cal{O}} = \tau N^{2(1-d)/(2-\gamma)} .
\ee
In the double scaling limit with fixing these scaling variables finite,
the singular part of the free energy becomes
\ba
F(t,t_{\cal{O}})
& = & -\frac{t^{2-\gamma}}{(2-\gamma)(1-\gamma)} - \bar{a}_1\log t
      - \sum_{h=2}^\infty\bar{a}_ht^{(1-h)(2-\gamma)} \nonu
&   & - \sum_{h=0}^\infty\sum_{k=1}^\infty\bar{b}_{h,k}
      t^{(2-\gamma)(1-h)}\left(t_{\cal{O}}t^{d-1}\right)^k .
\label{ScalingPP}
\ea
This scaling function depends only on the scaling variables $t$ and
$t_{\cal{O}}$ and is the sum over the contributions of the surfaces with
different topologies.
When we take the double scaling limit, we discard the analytic part.
Therefore its explicit form is not necessary in the conventional matrix
models.
However, when we modify the matrix models by square terms, this non-singular
part becomes very important as discussed in \S\ref{MMMod}.

\vspace{3mm}
%
%

%
\end{document}